\documentstyle[prl,aps,preprint]{revtex}
\begin{document}
\draft
\title{Search for Charged Strange Quark Matter Produced in 11.5 A GeV/c
Au + Pb Collisions}
\maketitle
\begin{center}
\bigskip
\mbox{T.A. Armstrong                \unskip,$^{7}$}
\mbox{K.N. Barish                   \unskip,$^{12,\ast}$}
\mbox{S.J. Bennett                  \unskip,$^{9}$}
\mbox{A. Chikanian                  \unskip,$^{12}$}
\mbox{S.D. Coe                      \unskip,$^{12}$}
\mbox{T.M. Cormier                  \unskip,$^{11}$}
\mbox{R. Davies                     \unskip,$^{8}$}
\mbox{G.De Cataldo                  \unskip,$^{1}$}
\mbox{P. Dee                        \unskip,$^{11,\dag}$}
\mbox{G.E. Diebold                  \unskip,$^{12}$}
\mbox{C.B. Dover                    \unskip,$^{2,\ddag}$}
\mbox{P. Fachini                    \unskip,$^{11}$}
\mbox{L.E.Finch                     \unskip,$^{12}$}
\mbox{N.K. George                   \unskip,$^{12}$}
\mbox{N. Giglietto                  \unskip,$^{1}$}
\mbox{S.V. Greene                   \unskip,$^{10}$}
\mbox{P. Haridas                    \unskip,$^{6}$}
\mbox{J.C. Hill                     \unskip,$^{4}$}
\mbox{A.S. Hirsch                   \unskip,$^{8}$}
\mbox{R.A. Hoversten                \unskip,$^{4}$}
\mbox{H.Z. Huang                    \unskip,$^{3,\S}$}
\mbox{B. Kim                        \unskip,$^{11}$}
\mbox{B.S. Kumar                    \unskip,$^{12,\|}$}
\mbox{T. Lainis                     \unskip,$^{9}$}
\mbox{J.G. Lajoie                   \unskip,$^{12}$}
\mbox{R.A. Lewis                    \unskip,$^{7}$}
\mbox{Q. Li                         \unskip,$^{11}$}
\mbox{B. Libby                      \unskip,$^{4,\P}$}
\mbox{R.D. Majka                    \unskip,$^{12}$}
\mbox{M.G. Munhoz                   \unskip,$^{11}$}
\mbox{J.L. Nagle                    \unskip,$^{12,\ast\ast}$}
\mbox{I.A. Pless                    \unskip,$^{6}$}
\mbox{J.K. Pope                     \unskip,$^{12}$}
\mbox{N.T. Porile                   \unskip,$^{8}$}
\mbox{C.A. Pruneau                  \unskip,$^{11}$}
\mbox{M.S.Z. Rabin                  \unskip,$^{5}$}
\mbox{A. Raino                      \unskip,$^{1}$}
\mbox{J.D. Reid                     \unskip,$^{7,\dag\dag}$}
\mbox{A. Rimai                      \unskip,$^{8,\ddag\ddag}$}
\mbox{F.S. Rotondo                  \unskip,$^{12}$}
\mbox{J. Sandweiss                  \unskip,$^{12}$}
\mbox{R.P. Scharenberg              \unskip,$^{8}$}
\mbox{A.J. Slaughter                \unskip,$^{12}$}
\mbox{G.A. Smith                    \unskip,$^{7}$}
\mbox{P. Spinelli                   \unskip,$^{1}$}
\mbox{B.K. Srivastava               \unskip,$^{8}$}
\mbox{M.L. Tincknell                \unskip,$^{8}$}
\mbox{W.S. Toothacker               \unskip,$^{7}$}
\mbox{G. Van Buren                  \unskip,$^{6}$}
\mbox{W.K. Wilson                   \unskip,$^{11}$}
\mbox{F.K. Wohn                     \unskip,$^{4}$}
\mbox{E.J. Wolin                    \unskip,$^{12,\S\S}$}
\mbox{Z. Xu			    \unskip,$^{12}$}
\mbox{K. Zhao                       \unskip$^{11}$}
\it
\vskip \baselineskip
\centerline{(The E864 Collaboration)}
  $^{1}$
     University of Bari/INFN, Bari, Italy \break
  $^{2}$
     Brookhaven National Laboratory,
     Upton, New York 11973 \break
  $^{3}$
     University of California at Los Angeles,
     Los Angeles, California 90095 \break 
  $^{4}$
     Iowa State University,
     Ames, Iowa 50011 \break
  $^{5}$
     University of Massachusetts,
     Amherst, Massachusetts 01003 \break
  $^{6}$
     Massachusetts Institute of Technology,
     Cambridge, Massachusetts 02139 \break
  $^{7}$
     Pennsylvania State University,
     University Park, Pennsylvania 16802 \break
  $^{8}$
     Purdue University,
     West Lafayette, Indiana 47907 \break
  $^{9}$
     United States Military Academy,
     West Point, NY 10996 \break
  $^{10}$
     Vanderbilt University,
     Nashville, Tennessee 37235 \break
  $^{11}$
     Wayne State University,
     Detroit, Michigan 48201 \break
  $^{12}$
     Yale University,
     New Haven, Connecticut 06520 \break
\end{center}
\begin{abstract}
We present results of a search for strange quark matter (strangelets) in
11.5~A~GeV/c Au+Pb collisions from the 1994 and 1995 runs of experiment E864
at Brookhaven's AGS.  We observe no strangelet candidates and set a 90\%
confidence level upper limit of approximately $3\times 10^{-8}$ per 10\%
central interaction for the production of $|Z|=1$ and $|Z|=2$ strangelets
over a large mass range and with metastable lifetimes of about 50~ns or more.
These results place constraints primarily on quark-gluon plasma based
production models for strangelets.
\end{abstract}
\pacs{25.75.-q}
\narrowtext 

Color-singlet hadrons with baryon number $A>1$, called {\it quark matter}, are
allowed in the Standard Model but have never been observed by experiment.
All the quarks within this type of state would be free within the hadron's
boundary, and would not be subject to grouping into the familiar
$A=1$ baryons.  In this way it is different from a nuclear state,
which is a conglomerate of $A=1$ baryons.

Quark matter states containing up and down quarks, if they exist, are less
stable (more massive) than nuclei with the same baryon number and
charge, since nuclei do not decay into quark
matter.  This is presently understood to be a consequence of the relatively
large Fermi energy of two-flavor quark matter.  However, additional quark
flavors could possibly reduce the Fermi energy of quark matter\cite{fermigas}.
Hence strange quark matter (SQM),
which would contain strange quarks in
addition to up and down quarks, might be more stable than non-strange quark
matter with the same $A$, despite the mass of the strange quark.  Other
quarks are usually not considered since they are much more massive than the
strange quark, and thus are not expected to enhance stability.  Since SQM
systems are expected to contain approximately equal numbers of up, down,
and strange quarks (with
charges +2/3$e$, -1/3$e$, and -1/3$e$, respectively), they would
have lower charge-to-mass ratios than nearly all ordinary nuclei.  This
property is the basis for all current SQM searches at heavy ion accelerators.

Studies have used quantum chromodynamics (QCD) and the MIT Bag Model of
hadrons~\cite{chodos} to treat SQM
quantitatively~\cite{chin_kerman,farhi_jaffe,gilson_jaffe}.  All of the
theories contain the feature that SQM systems become more stable as $A$
increases, due to the small total charge of SQM as well as bag model
effects.  For sufficiently large $A$, SQM may be absolutely
stable~\cite{witten}.  For smaller $A$, SQM may be metastable, that is
stable against strong decays but subject to weak decays with lifetimes in the
range $10^{-4}$ to $10^{-10}$ sec~\cite{chin_kerman,berger,dalitz}.
SQM systems with $A \le 100$, which might be produced in high energy
heavy ion collisions, are predicted to be metastable for a wide range of SQM
properties and bag model parameters~\cite{dalitz}.  These smaller systems are
commonly called {\it strangelets}.

Three types of production model have been applied to strangelet production in
nucleus-nucleus collisions.  In the first type, called
{\it coalescence models}, a group of known $A=1$
particles are made which, in sum, contain the same quantum numbers (baryon
number, strangeness, and charge) as a viable strangelet, and then these
ingredients fuse to form a strangelet~\cite{baltz}.
A second type of production, called {\it thermal models}, assume further
that chemical and thermal equilibrium are achieved
prior to final particle production~\cite{pbj}.  Coalescence and thermal models
usually predict lower strangelet cross sections than the last type of model,
in which an intermediate
quark-gluon plasma (QGP) state is formed after the initial nucleus-nucleus
collision, and the QGP loses energy in a way that possibly favors strangelet
production.  Kapusta {\it et al.} have estimated that a QGP would be produced
between 0.1\% and 1\% of central (small impact parameter) Au+Au collisions
at AGS energies~\cite{kapusta}.  Greiner {\it et al.} have suggested that
a large fraction of such QGP states would evolve into a strangelet
by a strangeness distillation mechanism~\cite{greiner}; other
distillation estimates predict a wide range of production
levels~\cite{liu,crawford}.  Thus strangelet production could be as high as
$10^{-4}$ to 10$^{-3}$ per central Au+Au collision, well within
the sensitivity to be presented here.  Note also that a
strangelet produced by the strangeness distillation of a QGP could have
approximately the same $A$ as the QGP itself, since the QGP would largely
lose energy by meson -- not baryon -- emission.  Hence it is of considerable
interest for experiments to remain sensitive to a large mass range.

	Early strangelet searches in Si+Cu collisions~\cite{barrette}
and in S+W collisions~\cite{borer} yielded null results.  More recently,
experiments utilizing Au beams at BNL\cite{e878,e886} and
Pb beams at CERN\cite{na52} saw no evidence for strangelet production, despite
the increased production potential of these heavier beams.
The experiments were sensitive to particles with proper lifetimes of about
50~ns or more, depending on the experiment.  All these
experiments, with the exception of the one described in
Ref.~\cite{barrette}, used focussing
spectrometers which, at a given magnetic field setting, have good acceptance
for a fixed rigidity $R=p/Z$, where $p$ is the
momentum and $Z$ is the charge of the produced particle.  The production
limits obtained using these spectrometers are strongly dependent upon the
production model assumed for
high mass particles such as strangelets.  In this paper we show the results
of an open geometry spectrometer experiment (containing dipole-type magnets
only)
whose sensitivity is less subject to the shape of a
particle's differential cross section.  We examine the mass range
$m \ge 5$~GeV/c$^2$ and $m \ge 6$~GeV/c$^2$ for $Z=+1$ and $Z=+2$ respectively,
and $m \ge 5$~GeV/c$^2$ for $Z=-1$ and $Z=-2$.  We are sensitive to particles
with proper lifetimes greater than about 50~ns.

A schematic diagram of the E864 spectrometer is shown in 
Fig.~\ref{fig:plan_elev}.  An 11.5 GeV/c per nucleon Au beam enters from the
left through a quartz Cerenkov beam counter and veto counters\cite{beam},
and is incident on a Pb target.  A segmented scintillator
multiplicity counter measures an interaction's products within an
angular range of 16.6$^\circ$ to 45$^\circ$ with respect to the
incident beam, providing a rough measure of the impact parameter, or
centrality, of the reaction~\cite{beam}.  For this analysis, we require the
multiplicity counter's pulse height to exceed a threshold such that 10\% of
the total Au+Pb cross section is accepted.  This multiplicity trigger
thus accepts the 10\% most central (smallest impact parameter) events.
Interaction products which are within the
experimental acceptance pass through two dipole magnets labelled M1 and M2, and
proceed through downstream detectors.  Three segmented planes of scintillation
counters (hodoscopes) labelled H1, H2, and H3, each contain 206 vertical
scintillator slats
viewed with photomultiplier tubes located at the top and bottom of the slats.
Each photomultiplier signal is digitized for both pulse height and time
information.  Three arrays of 4 mm diameter straw tubes labelled S1, S2 and S3,
provide high resolution position measurements.  The straw signals are
digitized in a latch system.  Each array includes three planes of doublet
layers. Two of the layers are inclined at $\pm 20^\circ$ to the
vertical, so that they provide a measurement of the vertical as well
as the horizontal coordinate.  S1 was not used in this analysis.
A lead/scintillating fiber hadronic calorimeter labelled CAL terminates the
apparatus~\cite{e864_cal}.  It consists of 754 towers, each of which is
read out by a photomultiplier tube.  These photomultiplier signals are
digitized for both pulse height and time information, and thus provide energy
and time-of-flight measurements.
The calorimeter is also used to form a high-level trigger that
correlates the energy and time-of-flight signature of showers on a
tower-by-tower basis, and is set to identify particles of high mass.  This
trigger, called the {\it late-energy trigger} or LET, provides the experiment
with a rejection factor of about 50 in 10\% central interactions.
A paper on the E864 apparatus is forthcoming~\cite{taa}.

The charge ($Z$) of a
particle that traverses the spectrometer is measured using pulse height
information from the 3 hodoscope walls.  Its rigidity ($R$) is derived from the
target position and downstream slope and position
of the particle's track in the spectrometer's magnetic bend plane, as
measured by the straw tube and hodoscope detectors.  The particle's velocity
is measured using timing information from the hodoscopes, and this gives the
relativistic quantities $\beta$ and $\gamma$.  The particle's
mass ($m$) is then reconstructed as $m = {R \over \gamma \beta}Z$.
The calorimeter's time and energy information was used to confirm the
above measurements or reject potential backgrounds.
This analysis was confined to a rapidity range about 1.3 units wide near
the center-of-mass rapidity value of 1.6, as we expect strangelet production
to be peaked in this region.

The strangelet analysis presented here uses over 120 million 10\% central Au+Pb
events taken from different magnetic field settings during 2 separate running
periods.  A preliminary strangelet search for positively-charged strangelets
was performed in 1994 with a partially-completed apparatus.
Analysis methods were developed
largely using data from this first run, and the experiment's capabilities were
learned, especially concerning the dominant background process in our
spectrometer~\cite{barish,barish2,rotondo}.
Our spectrometer was completed and optimized for both
positively- and negatively-charged strangelet states in our 1995
run~\cite{nagle,coe}, and these searches achieved excellent sensitivity due
to the high-rejection LET trigger.

An example mass distribution derived from our 1995 run for $Z=2$ particles is
shown in Fig.~\ref{fig:mass_dist}.  All tracking and calorimeter cuts were used
to produce this plot, but no corrections for efficiency or geometrical
acceptance were applied.  Peaks for $^3$He, $^4$He, and $^6$He show
prominently in the
figure.  The mass resolutions obtained for these states and others such
as $p$, $d$, $t$, $K^{-}$, and $\bar{p}$ are as expected
considering the detector resolutions and multiple scattering in the
spectrometer.

We observe no strangelet candidates in our 1995 data with $m>5$~GeV/c$^2$ for 
$Z=+1$, $Z=-1$, and $Z=-2$ systems, and we observe no candidates with
$m>6$~GeV/c$^2$ for $Z=+2$ systems.

In order to set limits on strangelet production, we compute the following
expression for the number of candidates observed:
\begin{equation}
  N_{obs} = \frac{I_{central}}{\sigma_{central}} \int \epsilon(y,p_{t})
        \frac{d^{2}\sigma}{dy dp_{t}} dy dp_{t},
\label{eq:nobs}
\end{equation}
where $N_{obs}$ is the number of strangelets observed, $I_{central}$ is the
number of central interactions examined, $\sigma_{central}$ is the cross
section for 10\% central Au+Pb interactions (10\% of the total Au+Pb cross
section),
$\epsilon(y,p_{t})$ is the
efficiency for detecting a strangelet as a function of $y$ and transverse
momentum ($p_{t}$),
and $ d^{2}\sigma / dydp_{t} $ is the strangelet differential cross
section.  We take the differential cross section to be separable in $y$ and
$p_{t}$:
\begin{equation}
        \frac{ d^{2} \sigma }{ dy dp_{t} } = \sigma_{s}
 \left[ \left( \frac{2}{ <p_{t}> } \right)^{2} p_{t}
e^{ \frac {-2p_{t}} {<p_{t}>} } \right]
 \left[ \frac{1}{\sqrt{2\pi}w} e^{\frac{-(y-y_{cm})^{2}}{2w^{2}}} \right] ,
\label{eq:diffxs}
\end{equation}
where $\sigma_{s}$ is the total strangelet cross section in central collisions,
$y_{cm}$ is the center-of-mass rapidity, and $w$ is the RMS width
(standard deviation) of the rapidity distribution of the strangelet.  We take
$w=0.5$ and $<p_t>$ = $0.6\sqrt{A}$~GeV/c.

Since $N_{obs}=0$ in our analysis, we can say from Poisson statistics
that there is a 90\% chance that $N_{obs}<2.3$.  By inverting Eq.~\ref{eq:nobs},
we obtain a 90\% confidence level (90\% C.L.) upper limit on strangelet
production per central interaction.  Figure~\ref{fig:limits} show E864's
90\% C.L. limits for positive and negative strangelets with lifetimes greater
than 50~ns produced in 11.5 GeV/c Au+Pb interactions.  The 4 curves which
display our 1995 results in Fig.~\ref{fig:limits} begin well
above the mass distributions of known particles reconstructed in
our data.  These starting values are 4.7, 4.7, 5.6, and 7.5 GeV/c$^{2}$ for
$Z=-2,-1,+1,+2$, respectively.
Also shown in the figure are our 1994 results, which are more fully
described in Refs.~\cite{barish,rotondo}.  

	E864's upper limits are nearly flat as a function of mass, owing to the
large acceptance of the spectrometer.  These limits are only mildly sensitive
to changes in
Eq.~\ref{eq:diffxs} for the same reason.  For example, if the rapidity width of
strangelet production were taken to be $w = 0.5/\sqrt{A}$, the E864
curves in Fig.~\ref{fig:limits} would be lower (give better limits) by less
than a factor of 2, while this change could have a strong, adverse effect on
a focussing spectrometer experiment.

	E864's upper limits constrain either the properties of SQM or the
cross section for its production in heavy ion interactions.  Two SQM
properties in particular may be restricted.  First, SQM lifetimes may
be constrained to $\tau << 50$~ns.  This possibility is unlikely for a wide
range of strangelets which are expected to undergo semileptonic and radiative
decays only~\cite{dalitz}, as lifetime estimates are based on
accepted quantities such as phase space, vertex suppression factors in
Feynman diagrams, and weak-decay lifetimes.  Second, our limits may constrain
bag model parameters as applied to SQM so that SQM is either
unstable for all $A$ or metastable only when $A>>100$.  This possibility,
while intriguing, must remain unanswered until the issues of strangelet
production are fully addressed.
	
	The sensitivity of this analysis is comparable to the coalescence
production levels for low-mass strangelets.  For example, a strangelet with
$A=7$ and strangeness $S=-4$ could be produced at approximately the same level
as the hypernucleus $^{7}_{\Xi^{0}\Lambda\Lambda}$He, since these states have
the same quantum numbers $A$ and $S$.  Ref.~\cite{baltz} estimates
$^{7}_{\Xi^{0}\Lambda\Lambda}$He will be produced between $3\times 10^{-8}$
and $7.2\times 10^{-8}$ per central Au+Au collision at the AGS, while our
sensitivity for this state is about $6\times 10^{-8}$ per central Au+Pb
collision.  It appears, however, that the model in Ref.~\cite{baltz} is
optimistic, since a preliminary analysis of data shows that the model
over-predicts light nucleus production~\cite{pope}.  Thermal models
would predict
production below our sensitivity for low mass strangelets.  For example,
the $^{7}_{\Xi^{0}\Lambda\Lambda}$He rate is computed to be
$\sim 2\times 10^{-10}$ in Au+Au collisions in Ref.~\cite{pbj}.  For larger
mass strangelets, both coalescence and thermal models predict production below
our sensitivity.  Our limits do constrain the sequence of QGP
production~\cite{kapusta} followed by QGP decay into a
strangelet~\cite{greiner}.
For a $10\le m \le 100$~GeV/c$^2$ strangelet with $|Z|=1$ or $|Z|=2$ and
lifetime above 50~ns, our data approximately
restricts these processes ({\it cf.} with Fig.~\ref{fig:limits}) at the 90\%
confidence level as follows:
\begin{equation}
{\rm BR}({\rm Au+Pb} \to {\rm QGP}) \times {\rm BR}({\rm QGP} \to
{\rm Strangelet})\; ^{<}_{\sim} \; 3 \times 10^{-8},
\label{eq:z1lim}
\end{equation}
where BR(Au+Pb$\to$QGP) is the probability for a 10\% central Au+Pb collision
at 11.5 GeV/c to produce a QGP, and BR(QGP$\to$Strangelet) is the probability
of the QGP to decay into the strangelet in question.
Some QGP production estimates for strangelets are largely
ruled out by our results~\cite{liu}, while others are being challenged.
For example, Ref.~\cite{crawford} predicts a strangelet with $A=10$, $Z=2$
to be produced $7.5\times 10^{-8}$ per central Si+Au interaction (with expected
higher yields in Au+Pb interactions), which is
above our limit of $5.3\times 10^{-8}$ per central Au+Pb interaction for the
same strangelet.

	In summary, we have found no evidence for strangelet production in
11.5 GeV/c per nucleon Au+Pb collisions,
and set a 90\% confidence level upper limit of about $3 \times 10^{-8}$
per 10\% central Au+Pb interaction for the production of $Z=|1|$ and $Z=|2|$
strangelets over a wide mass range and with lifetimes about 50~ns or more.
This represents the highest sensitivity strangelet search yet achieved in a
heavy ion experiment at AGS energies.

We gratefully acknowledge the efforts of the BNL AGS staff.
This work was supported by grants from the U.S. Department of Energy's
High Energy and Nuclear Physics Divisions, the U.S. National
Science Foundation, and the Istituto Nationale di Fisica Nucleare of Italy.

\begin{figure}
   \caption{Schematic views of the E864 spectrometer. In the
      plan view, the downstream vacuum chamber is not shown. M1 and M2 are
      dipole analyzing magnets, S2 and S3 are straw tube arrays, H1-H3 are
      scintillator hodoscopes, and CAL is a hadronic calorimeter.  The
      horizontal and vertical scales are in meters.}
   \label{fig:plan_elev}
\end{figure}

\begin{figure}
   \caption{$Z=+2$ mass distribution for $1.1<y<2.2$. No correction for
	acceptance has been applied.}
   \label{fig:mass_dist}
\end{figure}

\begin{figure}
   \caption{90\% confidence level limits for $|Z|=1$ and $|Z|=2$ strangelet
production in 10\% central Au+Pb collisions, for strangelets with lifetimes
greater than 50 ns.  The solid lines correspond to $|Z|=1$, while the
dashed lines correspond to $|Z|=2$.}
   \label{fig:limits}
\end{figure}


\begin{references}
\bibitem{fermigas}	In the limit of a simple relativistic Fermi Gas of
			massless quarks, the energy of quark matter is
			proportional to $n_f^{-1/3}$, where $n_f$ is the number
			of quark flavors in the system.
\end{references}
\end{document}